\documentstyle[12pt,aaspp4,epsfig]{article}

\def\deg{\ifmmode^\circ\else$^\circ$\fi}

\def\mic{~$\mu$m}

\def\ie{{\it i.e.}}

\def\et{{\it et al.}}

\def\arcs{\ifmmode {''}\else $''$\fi}
\def\arcm{\ifmmode {'}\else $'$\fi}
\def\parcs{\sa=.07em \sb=.03em
     \ifmmode $\rlap{.}$^{\scriptscriptstyle\prime\kern -\sb\prime}$\kern -\sa$
     \else \rlap{.}$^{\scriptscriptstyle\prime\kern -\sb\prime}$\kern -\sa\fi}
\def\parcm{\sa=.08em \sb=.03em
     \ifmmode $\rlap{.}\kern\sa$^{\scriptscriptstyle\prime}$\kern-\sb$
     \else \rlap{.}\kern\sa$^{\scriptscriptstyle\prime}$\kern-\sb\fi}
\def\mpc3{\mbox{Mpc$^{-3}$}}
\def\Msun{M$_{\odot}$}

\def\kp{{\rm K}$^{\prime}$}
\def\lya{{\rm Ly}$\alpha$}

\def\han {\mbox{{\rm H}$\alpha$}}
\def\pa {\mbox{{\rm P}$\alpha$}}
\def\ha{\han}

\def\brg {{\rm Br}$\gamma$}

\def\spose#1{\hbox to 0pt{#1\hss}}
\def\simlt{\mathrel{\spose{\lower 3pt\hbox{$\mathchar"218$}}
     \raise 2.0pt\hbox{$\mathchar"13C$}}}
\def\simgt{\mathrel{\spose{\lower 3pt\hbox{$\mathchar"218$}}
     \raise 2.0pt\hbox{$\mathchar"13E$}}}
\def\lsim{\rlap{$<$}{\lower 1.0ex\hbox{$\sim$}}}
\def\gsim{\rlap{$>$}{\lower 1.0ex\hbox{$\sim$}}}

\begin{document}

\title{   A Narrowband Imaging Search for [OIII] Emission 
from Galaxies at z$>$3}

\author { Harry I. Teplitz }
\affil{NOAO Research Associate \\
Goddard Space Flight Center, Code 681, GSFC, Greenbelt, MD 20771 \\
hit@binary.gsfc.nasa.gov}

\author {Matthew A. Malkan, Ian S. McLean}
\affil{Department of Physics \& Astronomy, Division of Astronomy \\
University of California at Los Angeles \\
Los Angeles, CA  90095-1562 \\
 malkan@astro.ucla.edu, mclean@astro.ucla.edu}

\begin{abstract}
  
  We present the results of a narrow-band survey of QSO fields at
  redshifts that place the [OIII](5007\AA) emission line in the
  $\frac{\Delta\lambda}{\lambda} \sim$ 1\% 2.16\mic~filter.  We have
  observed 3 square arcminutes and detected one emission line
  candidate object in the field around PC 1109+4642.  We discuss the
  possibilities that this object is a star-forming galaxy at the QSO
  redshift, $z_{em}=3.313$ or a Seyfert galaxy.  In the former case,
  we infer a star formation rate of 170\Msun/yr for this \kp=21.3
  object.  The galaxy has a compact but resolved morphology, with a
  FWHM=0.6\arcs, or 4.2kpc at z=3.313 (H$_{0}=50$~km
  s$^{-1}$Mpc$^{-1}$~and q$_{0}=0.5$).  The comoving density of such
  objects in QSO environments appears to be 0.0033\mpc3, marginally
  lower ($\le 3\sigma$) than the density observed for \ha-emitters in
  absorption-line fields at z$\sim 2.5$, but similar to the density of
  Lyman Break Galaxies at z$\sim$3.  If on the other hand, most of the
  line emission is [OIII] from a Seyfert 2 nucleus at z=3.31, then the
  high inferred volume density could imply a large evolution in the
  Seyfert 2 luminosity function from the current epoch.  We find the
  field containing the object to also contain many faint extended
  objects in the \kp~image, but little significant excess over the
  expected number-magnitude relation.  We discuss the implication of
  the emission line being a longer wavelength line at a lower
  redshift.

\end{abstract}


\section {Introduction}

High redshift galaxy surveys were, once limited to detection of
quasars and radio-loud AGN, have recently found a large population of
normal galaxies at z$>$2.  Steidel et al. (1996, 1998) have found
star-forming ``Lyman break galaxies'' (LBG) to be common at $z>3$,
greatly increasing the current understanding of the evolution of
normal galaxies.  In addition, deep narrow-band imaging has been used
in several successful \lya~searches (Cowie \et~1998, Hu \& McMahon
1996, Francis \et~1996).  Though optical searches have the advantage
of large area coverage, they are limited by the effects of dust
reddening and a selection bias for the youngest (bluest) objects.  The
near-IR continuum, by contrast, samples the older stars, giving an
independent view of the dominant stellar population and allowing
direct comparison with similar observations at other epochs.  At large
redshifts, the strong emission lines associated with on-going star
formation shift into the near infrared, allowing narrow-band imaging
surveys to identify starforming galaxies.

Given that small amounts of dust reddening can completely extinguish
\lya-emission, \ha~is the strongest emission line in many starburst
galaxies.  To date there have been numerous successes in detecting
emission line galaxies in the IR, targeted at \ha.  Given the
relatively small area covered by the current generation of
256$\times$256 pixel IR arrays, IR searches have typically centered on
known ``signposts'' such as quasars or absorbers where objects at the
target redshift are already known.  In Teplitz, Malkan \& McLean
(1998, hereafter TMM98), we detected 11 \ha-emitters in 12 square
arcminutes (down to a $1\sigma$~limiting line flux of $2\times
10^{-17}$ergs/cm$^2$/s), in fields centered on metal absorption-line
systems in quasar lines of sight.  The comoving space density of these
objects was found to be 0.0135$^{+0.0055}_{-0.0035}$~Mpc$^{-3}$, Such
a volume density is 3--5 times higher than the density of comparably
luminous Lyman break galaxies (LBGs) at similar redshifts.  The
average inferred SFR for the galaxies is 50 \Msun /yr, consistent with
other dereddened estimates for galaxies at z$>$2 (Dickinson, 1998).  A
similar search was conducted by Pahre \& Djorgovski (1996) with the
same instrument, though no detections were found in four fields, two
of which were targeted on \ha, one on [OIII] and one on [OII].
Thompson \et~(1996) covered $\sim$300 square arcminutes down to a
survey depth of $10^{-16}$ergs/cm$^2$/s.  In this area they discovered
a single strong line emitter, which they identify as a star forming
galaxy at z=2.43, with an inferred SFR$\ge$240\Msun/year, assuming it
is H$\alpha$ (Beckwith \et~1998).  In a companion survey targeted to
Damped \lya~and metal absorption-line systems, the same group found 18
targets in 163$\Box$\arcm, at comparable depth (Mannucci \et~1998,
hereafter MTBW98).

The [OIII] doublet is, in many cases, the other strongest emission
line. In fact, the [OIII]:\ha~ratio is about 0.6 in local star-forming
galaxies (Thompson, Djorgovski, \& Beckwith 1994 and references
therein).  The relative strength of [OIII] is greater in more
metal-poor gas, because the lower cooling efficiency results in a
higher level of ionization.  The primary complication of using [OIII]
as a tracer of star-formation is the tendency of active galactic
nuclei (AGN) also to produce strong [OIII] lines, powered by the hard
non-stellar ionizing continuum in the nucleus.  If a significant
portion of the [OIII] flux comes from the Narrow Line Region (NLR) of
an AGN, then this line would give an over-estimate of the
star-formation rate (Kennicutt 1992).  This risk is greatest for AGN
with pure NLR spectra, \ie, the Seyfert 2's, because their
[OIII]/H$\alpha$ ratios are typically three times higher than in
starburst galaxies.  However, this sensitivity to narrow-line AGN is
partially offset by the fact that they are rare.  Optical searches
(which observed the rest-frame UV emission lines) show the space
density of AGN is considerably smaller than that of star-forming
galaxies.  For example, Lyman Break Galaxies (Steidel \et 1996;
hereafter LBG) are detected at ten times the rate of AGN using the
Lyman dropout technique.  TMM98 presented arguments that not all of
the \han~flux in their galaxies is the result of active nuclei,
including the fact that their volume density would be too high.  We
will consider this issue further below when interpreting our
observational results.

In this paper we present the results of a narrow-band search for 
[OIII] emission from galaxies at z$>$3 in quasar environments.  
By extending the search for emission lines in the infrared, we can answer
two important questions.  First, we evaluate the usefulness of [OIII] as
a tracer of star formation at high redshift, and the prospects for using
this line in a larger survey.  An infrared search at z$>$3 will allow direct
comparison of the near IR techniques with the Lyman break method, finally
unifying the different approaches, and potentially applying each to the same
field (for a \lya~counterpart see Cowie \et~1998).  Secondly, we can use
observations pointed at signposts at z$>$3 to limit the {\it field}~density of
\ha-emitting objects near lower redshift signposts.  Previous targeted
searches have called for more extensive control fields.

In section 2 we present a summary of the observations, data reduction
and sensitivity limits.  In section 3 we present the results of the
narrow-band and continuum imaging.  In section 4 we discuss the
implications of the one survey candidate and the non-detections in
other fields.  Throughout the paper we assume H$_{0}=50$~km
s$^{-1}$Mpc$^{-1}$~and q$_{0}=0.5$.

\section{Observations}
    
Using the NASA/IPAC Extragalactic Database (NED) we selected fields
containing quasars at redshifts which would place redshifted [OIII]
($\lambda_{0} = 5007$\AA) in the \brg (2.16\mic) filter (see TMM98 for a
discussion of filter selection).  We required the redshifts to fall
within $3.299<z<3.329$.  All observations were taken with the Near IR
Camera (NIRC) on the 10m Keck I telescope; the camera has a
256$\times$256 pixel InSb array, (Matthews \et, 1994) and a field
of view of 38\arcs $\times$ 38\arcs; 0.15\arcs/pixel.  Table 1 lists
the fields observed, integration times and limiting sensitivities.  We also
list the FWHM of the seeing disk for each field as measured from the
quasar or other bright, unresolved objects in the field.  For
comparison, we include in Table 1 the field around QSO1159+123
surveyed by Pahre \& Djorgovski (1996) with the same instrument (in a
different filter).
   
Observations were taken and reduced with the same procedure described
in TMM98, which will only be summarized here.  We obtained images in a
sequence of ``dithered'' exposures, offsetting the telescope between
exposures in a 3$\times$3 grid spaced by 3\arcs.  The data were
reduced by dividing by a twilight flat and then subtracting a running
median sky frame created from the nine exposures taken closest in time
to each image.  Objects were identified using the SExtractor (Bertin
\& Arnouts 1996) software.  Photometry was performed using apertures
of 2.5 times the seeing disk.  The same aperture was applied to broad
and narrow-band exposures.  Photometric errors were estimated from
aperture photometry performed on random positions in the frame.
Errors in the narrow-minus-broad band color were estimated from Monte
Carlo simulations.  These simulations generated narrow and broad band
magnitudes for line-free objects having the gaussian errors measured
in the real data.  

\section{Results}

Figure 1 shows the narrow-minus-broad band color-magnitude diagram for
all the objects in the survey.  One object in the PC1109+4642 field
shows up as a clear candidate for excess flux in the narrow-band
filter.  The varying depths of the different fields makes the
significance of this measurement somewhat difficult to distinguish from
the combined figure, so we have not plotted the characteristic errors.
Instead, we have re-plotted the objects in the 1109 field separately
in Figure 2 along with the 99.5\%~confidence limits.
The 1109 field was observed on a night of
exceptionally good seeing, with a long integration time, and thus is more
sensitive than most of the other fields (see Table 1).  
The emission-line object was observed again, on a separate
night, to confirm the result and search adjacent sky area (though
conditions were substantially worse on the second night).  Good
agreement was seen in the \brg~magnitudes across the nights.  The
broad-band magnitude was not checked as thoroughly (as the object fell
in the low signal to noise ratio part of the dither pattern) and so
there is a discrepancy in the photometry from the two nights.  The
range between them is plotted as the error bars in the figure.  Figure
3 shows the 1109 field, together with the broad- and narrow-band
images of the candidate galaxy (hereafter 1109A).

We have also examined the continuum photometry for objects in each
field.  The 1109 field appears relatively crowded, but as already
noted it is a deeper exposure.  81 galaxies were measured in \kp~in
all the fields combined, of which 23 are in the 1109 field.  To
examine the significance of large number of galaxies, we have
calculated the number-magnitude relation for that field and for the
other five fields.  Figure 4 shows the number counts, also compared
with results from the literature.  No attempt was made at star-galaxy
separation for objects fainter than \kp=20, but fortunately at this
high galactic latitude, virtually all of these faint objects are
galaxies.  Completeness was estimated from recovery of mock objects in
similar data.  The number-magnitude relation for the other (four)
comparably deep fields match the expectation from K-band surveys in
the literature.  The number of excess \kp$\ge20$~galaxies in the 1109
field appears marginally significant (at the $\sim 2\sigma$~level).

In the broad-band image, 1109A appears extended, though its faint
magnitude makes its half-light radius hard to measure reliably (IRAF's
IMEXAM task gives a FWHM=0.6\arcs). In the \brg~image, the object has
FWHM=0.55\arcs.  The question arises whether the narrowband excess
emission is likely to be from an active nucleus.  This is possible,
though there is some evidence to the contrary.  The seeing disk is
measured to be 0.45\arcs~from the (brighter) QSO.  1109A has a
magnitude of \kp=21.3 and a broad-minus-narrow band color excess
($\Delta m$) of 1.5 magnitudes, relative to featureless objects.

Object 1109A appears to be within 1.3\arcs~of another galaxy.  However
only 1109A is strongly detected in the narrowband.  It is interesting
to note that two other high-z emission line galaxies have been
reported recently that also appear to be very close to another object.
Francis \et (1998) find that one of the AGN in the 2139-4344 cluster
(see Francis \et~1996, 1997)lies with 0.7\arcs of another faint red
galaxy in high-resolution NICMOS imaging.  They suggest that the two
galaxies are in the process of merging and that some of the extended
\lya~emission from which the object was initially identified results
from star formation induced by the merger.  Similarly, Beckwith
\et~(1998) report that cK39 (an \ha-emitting galaxy at z=2.43 from the
Thompson \et~(1996) survey) lies within 1.3\arcs~of another red
galaxy.  They, too, interpret these as merging objects.  We note that
in the case of the cK39 there are arguments favoring an AGN source for
the detected emission line.

\section{Interpretation}

There are three possible explanations for the apparent excess flux in
the \brg~filter.  The object could be at the targeted redshift of
the QSO.  Secondly the emission line could be \ha, the strongest
optical emission line associated with ongoing star formation.  Lastly,
it could be a rest-frame near infrared line at a much lower redshift.
The properties associated with the galaxy for each assumed emission line
are given in Table 2. 

\subsection{The [OIII] Emission Line at $z=3.31$}

The first, and we argue most likely explanation is that the flux is
due to the [OIII] $\lambda 5007$~ emission line at the targeted
redshift of 3.31.  If that were the case, the $K^{\prime}\simeq
21.3$~apparent magnitude of the object (0.0025$\mu$Jy) would
correspond to an absolute magnitude of $M_B\sim -24.1$.  Assuming a 1
Gyr burst of star formation at z$>$5, we find that the passive
evolution and K-correction from the GISSEL96 spectral synthesis models
of Bruzual \& Charlot (see Bruzual \& Charlot 1993) predict this
galaxy would be the precursor of a modern galaxy with 0.5$L_*$.  In
that case the galaxy's FWHM would correspond to 4.2kpc, in reasonable
agreement with the expectation for a galaxy at this redshift (see for
example Lowenthal \et~1996).  The projected separation of 1109A from
the QSO at z$_{em}$=3.313 would be 133kpc.

As discussed above, [OIII] is emitted both by starbursts as well as
the NLR in an AGN.  To distinguish unambiguously these two
possibilities, it will be necessary to obtain a spectrum in either the
optical or near-infrared.  Such observations are planned with the LRIS
(Oke \et~1995) and NIRSPEC (McLean \et~1998) instruments on the Keck
II telescope, and will be presented in a future paper.  Meanwhile,
however, we consider the implications of each possibility.

From the $\Delta m=1.5$~of broad-minus-narrow-band color, we obtain
EW=150\AA~in the rest frame.  Typical local spiral galaxies show a
range in $EW(\mbox{{\sc [OIII]}})=10-80$, when it is detected at all
(Kennicutt, 1992).  On the other hand, local galaxies typically have
SFRs a factor of 5--10 lower than that estimated for LBGs at $z>3$,
including extinction correction.  If the line emission is dominated by
current star formation, we can use Kennicutt's (1983) relation to
estimate the SFR based on [OIII] luminosity:

\begin{equation}
\mbox{SFR(total)} = \frac{L(\mbox{\sc{[OIII]}})}
     {6.7\times 10^{40}\mbox{ergs~s}^{-1}} 
     \mbox{\Msun} 
     \mbox{yr}^{-1}
\end{equation}
Under the assumption that the line is
[OIII] from H~II regions, its luminosity implies SFR=170\Msun/yr.

We can also calculate the implied space
density of emitters in QSO environments at that redshift.  Given our
rectangular search window, we survey 0.47Mpc$^3$~(physical volume) per
NIRC field, leading to a comoving number density of
$0.0033^{+.008}_{-.0027}$ \mpc3 (with a 3$\sigma$~upper limit on the
density of 0.026\mpc3, as estimated from Poisson statistics; see
Gehrels 1986).  This density can be compared to the 0.0135\mpc3 for
\ha-emitters at redshifts of 2.3--2.5 from TMM98.  We note that while
the TMM98 search was targeted to both QSO environments and
metal-absorption line system fields, most of the emitters detected
were at absorber redshifts.  A similar conclusion was reached in
MTBW98 (with a density of $9\times 10^{-4}$\mpc3, to shallower depth),
so it is perhaps not surprising that the density of objects in QSO
fields may be lower.

Another factor to consider is evolution in the density of objects with
redshift.  We can compare our inferred z=3.3 density to the density of
Lyman Break Galaxies at a similar redshift, which is 0.004\mpc3~at
2.75 (a redshift window of 2--3.5, see Madau \et~1997).  This
comparable density argues in favor of object 1109A being part of the
typical high-z population of galaxies.  Similarly, its SFR is not
unusually high for an (extinction corrected) LBG (see Pettini
\et~1997).  On the other hand, we must consider the second possible
source of [OIII] emission: a Seyfert nucleus.

\subsection*{Does the [OIII] Emission come from an AGN?}

In TMM98 we argue that the space density of \ha~emitters is
inconsistent with the assumption that all the candidate objects are
AGN.  So we can ask: Is the space density of [OIII] emitters inferred
from this one candidate also inconsistent with AGN?  To answer that
question, we must assume a luminosity function for Seyfert galaxies at
$z>3$.  We will extrapolate the QSO luminosity function of Warren
\et~(1994).  That function was defined for the rest-frame continuum
flux at 1216\AA, under the \lya~emission line.  We, of course, do not
have a measurement of the flux at that wavelength.  Instead, we will
make a rough estimate by assuming that the candidate object has the
same observed frame $V-$\kp$\simeq 4.8$~color as the (also
AGN-contaminated) MTM095355+545428, leading to a value of
$M_{C}=-21$~at $z=3.3$.  Using the evolving Schechter luminosity
function suggested by Warren et al.  leads to a comoving space density
of such objects $4\times 10^{-4}$ Mpc$^{-3}$.  This is almost an order
of magnitude lower than the density inferred from our one candidate,
but that is of course highly uncertain.  Thus the density of AGN does
not favor the interpretation of the possible [OIII] line having
non-stellar (Seyfert 1) origin, but it doesn't absolutely preclude it
either.

If 1109A has a Seyfert nucleus, it is more likely the object is a pure
narrow-line AGN (a Seyfert 2) than a broad-line object because the
former are relatively brighter by an order of magnitude in the
5007\AA~emission line.  (We will ignore the small additional
contribution from 4959\AA, especially because most of it would fall
outside our narrow-band filter).  Because the median equivalent width
is about 20\AA~in low-luminosity quasars (Boroson \& Green 1992) and
about 200\AA~in Seyfert 2 galaxies (as measured from the
spectrophotometry of De Bruyn and Sargent 1978), our search goes one
order of magnitude deeper down the Seyfert 2 luminosity function than
it does down the Seyfert 1/quasar luminosity function.  The observed
equivalent width would be extremely high for a quasar, but is within
the range of values seen in Seyfert 2's.  The Seyfert 2 hypothesis is also
far more consistent with the fact that the galaxy is spatially {\it
  resolved.}  (If a broad-line AGN actually were present, it would
probably make 1109A very blue in the rest-frame UV. For a typical
$\alpha=-0.5$, 1109A would have $I\sim 23$~which could be easily
checked with deep imaging.  None has yet been published.)

Even with a large amount of luminosity evolution it is hard to reconcile
the density of high-z line-emitting galaxies with the density of Seyfert 2's.
We must realize, however, that the density of Seyfert 2's is not well
known at high redshift.  Complete samples of Seyfert 2's have been
observed locally (Rush et al. 1993), and it is seen that Seyfert 1's and 2's
above $L_*$~have comparable densities.  So, if Seyfert 2's do not evolve
considerably more with redshift than Seyfert 1s, then the space density
of high luminosity Seyfert 2's would fall short of the density observed
in emission-line searches.

If 1109A is representative of such a numerous, strongly evolving
population of Seyfert galaxies (of either type 1 or 2), it might have
strong implications for their present-day descendants.  By raising the
volume density of AGN by an order-or-magnitude over what had been
known from color-searches for quasars, this density could imply that a
Seyfert phase was a common occurrence in the early evolution of most
currently normal galaxies.

Spectroscopic evidence will of course be vital in settling the issue,
but there are already some hints that this might be the case.  A
substantial proportion of the high-redshift galaxies discovered in
emission-line searches do in fact show {\it mixtures} of starburst and
Seyfert activity.  For example, we discovered an emission line galaxy
at z=2.495 in the environment of the multiple C IV absorber in the
line of sight of the quasar SBS0953+549 (Malkan, Teplitz \& McLean
1995, 1996).  Optical spectroscopy showed this galaxy to have young
stellar absorption features as well as very strong \lya~emission, and
weaker high-ionization emission lines which suggest an AGN
contribution ($\le 30$\%) to the continuum flux.  \lya~emission line
searches have also found AGN as well as star-forming galaxies.
\lya~has been seen as an indicator of both potentially high mass
clusters (Francis \et~1997,1998) and pre-merger proto-galactic clumps
(Pascarelle \et~1997), even though both contain a large number of AGN.
Thus whatever the source of excitation, since \lya~searches have
found large clusters of objects at high-z, perhaps with a surprising
proportion of active nuclei, it is reasonable to suppose that the less
extincted [OIII] line may do the same.

Finally, we must consider the possibility that the emission-line is
[OIII] but that the source of ionizing radiation is the nearby quasar,
not intrinsic star formation or nuclear activity.  At a minimum
separation of 133kpc, 1109A only intercepts 0.1\% of the emitting QSO
radiation.  It is unlikely that this fraction would be sufficient to
produce the strong [OIII] emission line in 1109A.  Assuming a typical
rest-frame equivalent width for [OIII] to be $\sim 80$\AA~in QSOs and
the underlying QSO continuum to be $1\times
10^{-17}$ergs/cm$^2$/s/\AA, gives an indication of the maximum
possible energy that could be responsible for the candidate object's
emission line.  This estimate agrees with the observed
narrow-minus-broad band color of the QSO.  The inferred energy,
however, could only account for less than 1\% of the observed line
flux in 1109A.

\subsection{The \ha~Emission Line at $z=2.3$}

A second possible explanation for the excess narrow-band flux is that
there is an \ha~emission line at z=2.3.  This assumption would lead to
an inferred luminosity $L\sim 0.3L_*$~based on the model discussed
above.  The inferred star formation from an \ha~line would be
SFR=45\Msun/yr, which is close to the average found by TMM98,
but lower than that in MTBW98 by a factor of 1.5. If the emission line
is \ha~then we calculate a comoving space density of
$0.0055^{+.009}_{-.004}$ \mpc3, a factor of 2.5 less than the TMM98
survey.  This again is a reasonable difference, given that at z=2.3,
the current survey is {\it un}targeted and so would be probing the
field galaxy population.  Thus, if this object is at z=2.3, we take it
to be a ($<3\sigma$) confirmation of the conclusions reached in TMM98
that metal absorber fields show a higher space density of star forming
galaxies.

\subsection{A Low Redshift Near-IR Emission Line}
  
The emission line could be something at a lower redshift, in the
rest-frame near-infrared.  The most likely transitions to consider are
\pa(1.875\mic), [FeII](1.64\mic), and [SIII](0.9532\mic).  In each
case, the galaxy would be considerably fainter, but would also have a
reasonably smaller inferred SFR.  To calculate the inferred SFR we
apply a standard line ratio to \ha~for each IR emission line and then
again refer to Kennicutt (1983).  We take \pa:\ha$=0.1$, which is the
value for Case B recombination (see for example Hill \et~1996 and the
references therein).  We assume [FeII]:\ha$=0.034$ (see Calzetti
1997).  Finally, we use the ratio [SIII]:\ha$=0.43$, which is the
unreddened ratio based on observations of Orion often used as a
diagnostic of extinction (see Waller \et~1988 and references therein).
While none of the inferred SFRs are surprisingly low for an assumed IR
emission line (see Table 2), the inferred luminosities are, as they
would require the object to be at most 0.1$L_*$.  

If the emission line is one of the near-IR lines longward of 1\mic,
the implied densities of objects would be anomalously high for field
galaxies (based, however on a small dataset).  If we are seeing \pa~at
z=0.152, then we find a comoving density of $\sim 0.8$\mpc3. For
comparison, the comoving density of objects brighter than 0.001L* at
the current epoch is 0.014\mpc3 (Loveday \et~1992).  For [FeII] at
z=0.317 we find a density of 0.12\mpc3.

For [SIII] at z=1.266, the implied density would be 0.01\mpc3.  We can
compare this to the field density of objects at this redshift.  We use
the luminosity function at $1<z<2$~measured by Sawicki \et~(1997), to
calculate a comoving density of 0.023\mpc3~for galaxies brighter than
0.1$L_*$.  Thus on the basis of density alone we cannot rule out the
interpretation of the emission line as [SIII]. To investigate further
the implications of this interpretation, we can consider the dataset
obtained in TMM98 as well as the current observations.  In that
survey, seven fields were observed with the same 2.16\mic~filter, to
similar depths.  In that area, 5 line emitters were detected, one of
which as been spectroscopically confirmed to be a Seyfert 1 at z=2.3
and a second object which is anomalously bright and morphologically
suggestive of a Seyfert.  Considering the other three objects as a
limiting case, we combine those data with the current survey and find
an upper limit on the density of [SIII]-emitters at z=1.266 of $\sim
0.02$\mpc3.  Combining this with the inferred luminosity of
0.1L$_*$~makes [SIII] the most reasonable of the infrared emission
line possibilities.  Again, we do not consider this interpretation the
most likely, however.  We also could consider the lack of absorption
systems in the QSO spectrum (Schneider \et~1994) to be an argument
against this galaxy being identified by the [SIII] emission line, but
the projected distance is outside the radius expected for an absorber.

In summary, we have surveyed 3 square arcminutes down to a
$1\sigma$~limiting flux of $3\times 10^{-17}$ergs/cm$^2$/s.  In that
area we find a single emission-line object in the field around
PC1109+4642, with an emission-line flux of 7$\times 10^{-17}$
ergs/cm$^2$/s and \kp=21.3.  If this object is a
star-forming galaxy at the the QSO redshift,
we calculate a comoving space density and inferred SFR
consistent with other searches.

\acknowledgements

We would like to thank the Observing Assistants and Instrument
Specialists at the Keck telescopes, especially B. Schaefer,R.
Quick,T.Stickle,T.Bida,R.Cambell,B.Goodrich, and W. Harrison.  This
research has made use of the NASA/IPAC Extragalactic Database (NED)
which is operated by the Jet Propulsion Laboratory, California
Institute of Technology, under contract with the National Aeronautics
and Space Administration.  Data presented herein were obtained at the
W.M. Keck Observatory, which is operated as a scientific partnership
among the California Institute of Technology, the University of
California and the National Aeronautics and Space Administration.  The
Observatory was made possible by the generous financial support of the
W.M. Keck Foundation.

\clearpage

\begin{deluxetable}{lllccccc}
\tablenum{1}
\tablecolumns{8}

\tablecaption{Targets}
\tablehead{
\colhead{Field}&
\colhead{z}&
\colhead{date}&
\colhead{$t_{\rm{K}^{\prime}}$}&
\colhead{$t_{n.b.}$}&
\colhead{$\rm{K}^{\prime}$}&
\colhead{seeing}&
\colhead{1$\sigma$~line flux}\\
\colhead{}&
\colhead{}&
\colhead{}&
\colhead{(s.)}&
\colhead{(s.)}&
\colhead{$5\sigma$}&
\colhead{(arcs.)}&
\colhead{(ergs cm$^{-1}$s$^{-1}$)}
}

\startdata

Q0234+013       & 3.30  &  01/14/98 & 540  & 3600 & 20.0 & 0.45 & 5.3e-17 \\
PC1109+4642     & 3.313 &  03/17/98 & 2160 & 8640 & 21.1 & 0.45   & 2.3e-17 \\
PC1109+4642 N\tablenotemark{1} & 3.13  & 03/18/98 &  2160 & 6480  & 21.0 &  0.6 
& 2.6e-17\\
PC1153+4751     & 3.323 &  03/18/98 & 1620 & 6480 & 21.1 & 0.45  & 2.5e-17\\
Q1410+096       & 3.317 &  04/16/98 & 1620 & 4320 & 20.8 &  0.6  & 4.0e-17\\
PC1542+4744     & 3.312 &  06/27/96 & 1620 & 6480 & 21.2  &  0.6 & 2.1e-17\\
PC1542+4744 E\tablenotemark{1}  & 3.312 & 03/18/98 & 1620 & 4320 & 21.2 & 0.6 & 
4.2e-17 \\
Q1159+123\tablenotemark{2} & 3.502 & 04/05/94 & \nodata & \nodata & \nodata & 
0.75 & 3.15e-17 
\enddata

\tablenotetext{1}{The fields of PC1109 and PC1542 were reobserved, offset by 3/4 
of the
frame in order to examine an object of interest, so there is a 25\% overlap 
between 
these double fields.}
\tablenotetext{2}{Data on Q1159+123 taken from Pahre \& Djorgovski 1995}

\end{deluxetable}

\clearpage

\begin{deluxetable}{llllllcl}
\tablenum{2}
\tablecolumns{8}

\tablecaption{Properties of 1109A for various assumed line identifications}
\tablehead{
\colhead{Line}&
\colhead{z}&
\colhead{$M_{AB}$}&
\colhead{Lum.}&
\colhead{SFR}&
\colhead{FWHM}&
\colhead{Dist. to QSO} &
\colhead{density}\\
\colhead{}&
\colhead{}&
\colhead{}&
\colhead{ (L$_*$)}&
\colhead{M$_{\odot}$/yr}&
\colhead{(kpc)}&
\colhead{(kpc, projected)}&
\colhead{(\mpc3)}
}

\startdata

[OIII]       & 3.313  &  -24.3 & $>$0.5     & 171  & 4.2 & 133 & 0.0033 \nl
\ha          & 2.30   &  -23.5 & 0.3        & 45  & 4.7 & 150 & 0.0055 \nl
[SiIII]      & 1.27   &  -22.0 & 0.1        & 28  & 4.7 & 163 & 0.010  \nl
[FeII]       & 0.32   &  -18.8 & $<$0.01    & 18   & 2.4 & 108 & 0.12   \nl
\pa          & 0.15   &  -17.0 & $<$0.001   & 12 & 1.9 & 64  & 0.80   
\enddata

\end{deluxetable}

\clearpage

\figcaption[]{(\kp - 2.16\mic) vs. \kp~for all the fields surveyed.
  The + symbols are for objects in the 1153 field; the * for objects
  in the 1109North field.  The diamonds for objects in the 1109 field;
  the triangles for objects in the 1542 field; the squares for objects
  in the 1542east field; and the x for objects in the 1410 field.  The
  few objects that were observed twice between the 1109 and 1109N
  fields are plotted with the average value with error bars extended
  to the individual values.  The large error bar is the result of the
  low signal to noise re-observation of the emission-line object
  1109A}

\figcaption[]{(\kp - 2.16\mic) vs. \kp~for the 1109+4642 field.  Objects 
  that were observed twice between 1109 and 1109N are plotted as the
  average with error bars extended to the individual values.  The
  solid lines denote the three sigma errors for the 1109 field, not
  the 1109N field (which are larger).  The * indicate objects in the
  1109 field and the + indicate objects in 1109N.  1109N was observed
  in poorer conditions, so the two + symbols outside the errors are
  not actually 3 sigma detections.  The large error bar is the result
  of the low signal to noise re-observation of the emission-line
  object 1109A}

\figcaption[]{The \kp~image of the 1109+4642 field.  The small images
  at the bottom compare the broad-band (left) and narrow-band (right)
  observations of the line emitter.  The region compared is indicated
  by the square in the larger picture.}

\figcaption[]{\kp~number counts for objects in the 1109 field compared
  to objects in the other fields surveyd.  The squares with error bars
  indicate the counts for all the other fields, while the asterisks
  with error bars indicate the counts in the 1109 field.  Open symbols
  show the counts with no completeness correction.  The crosses show
  the number counts from the literature (as compiled by Gardner
  1998).}

\clearpage

\begin{figure}
\plotone{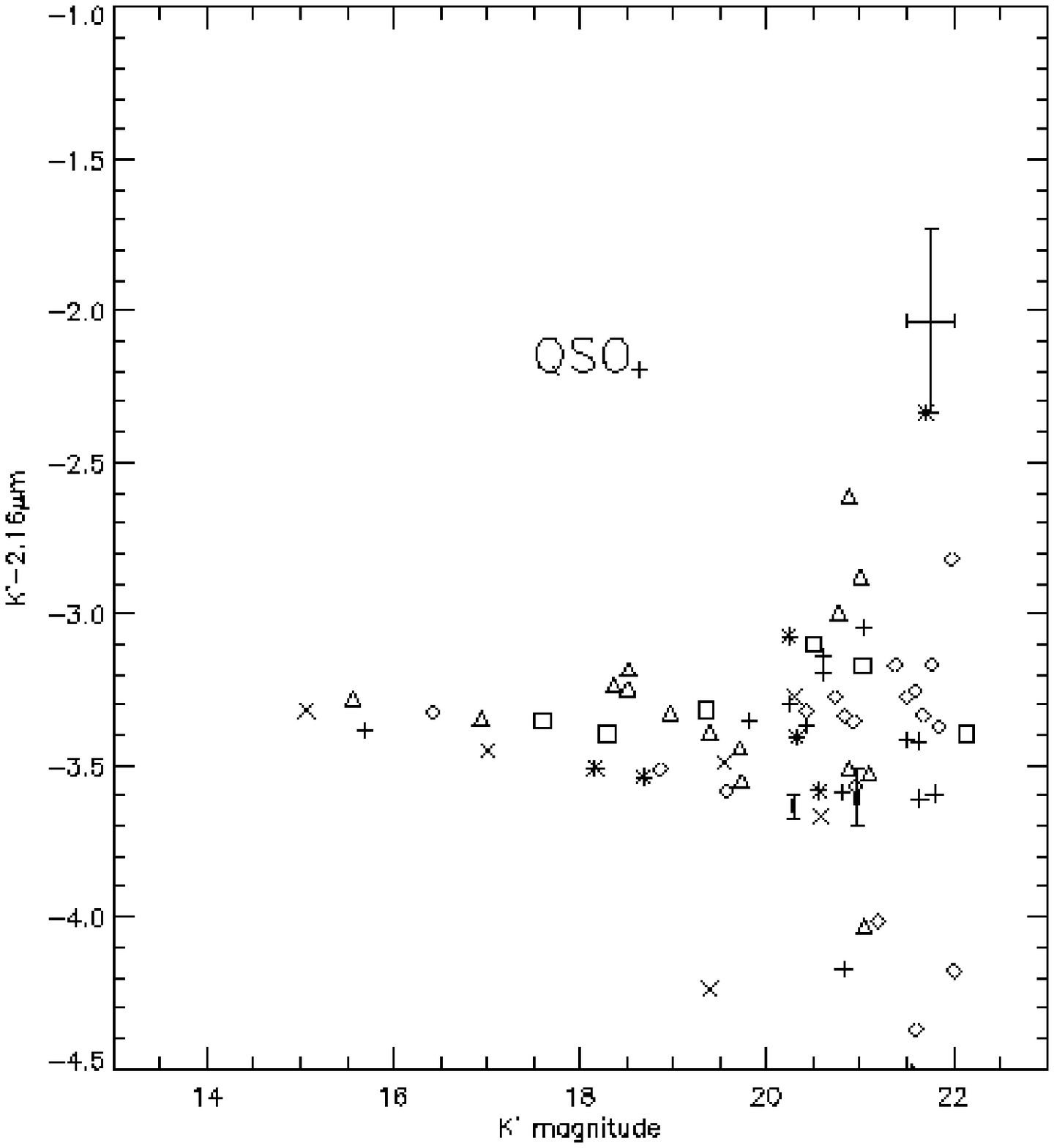}
\end{figure}

\clearpage

\begin{figure}
\plotone{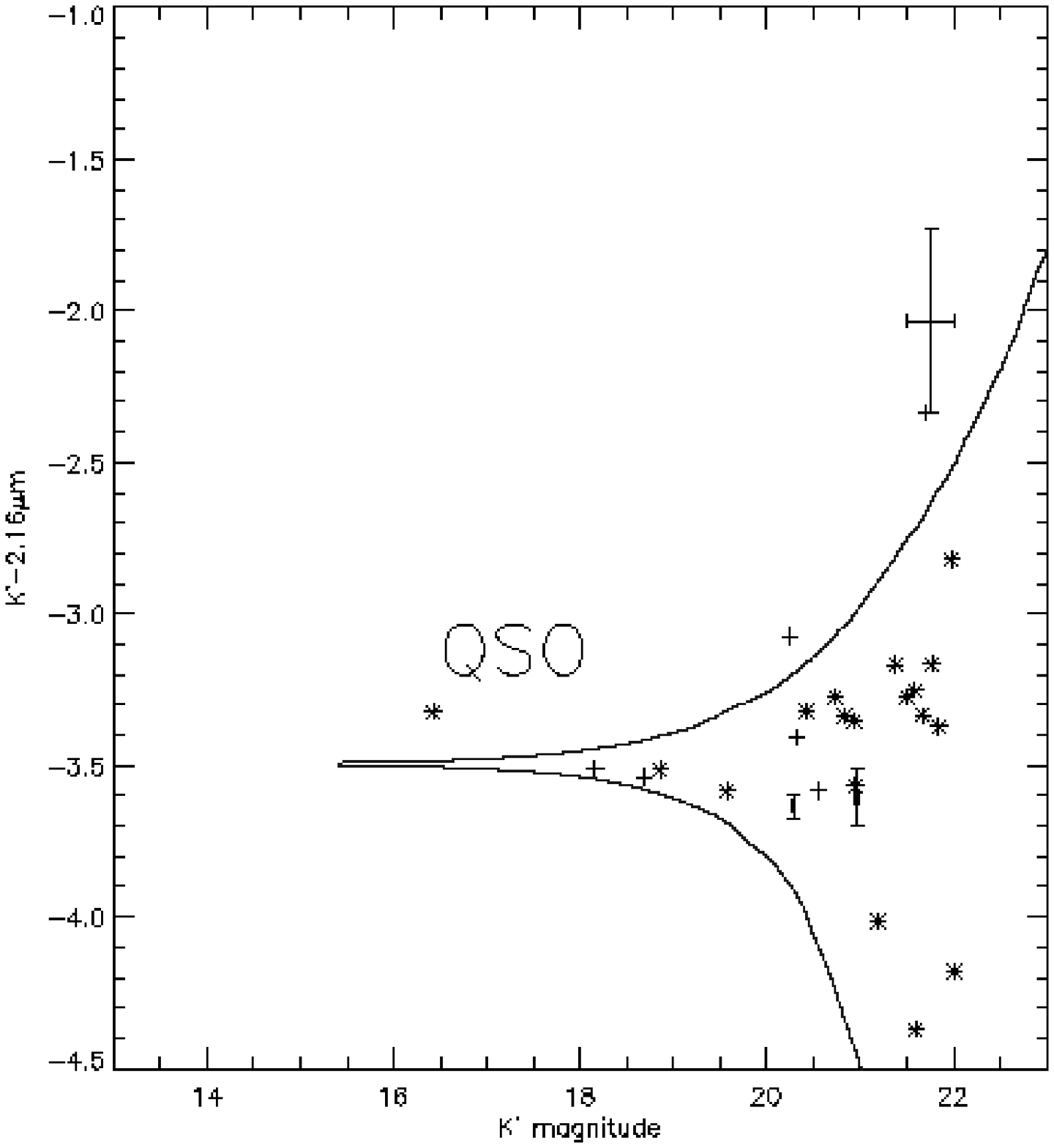}
\end{figure}

\clearpage

\begin{figure}[t]
\vskip -0.5in
\parbox{6in}{\epsfysize=6in \epsfbox{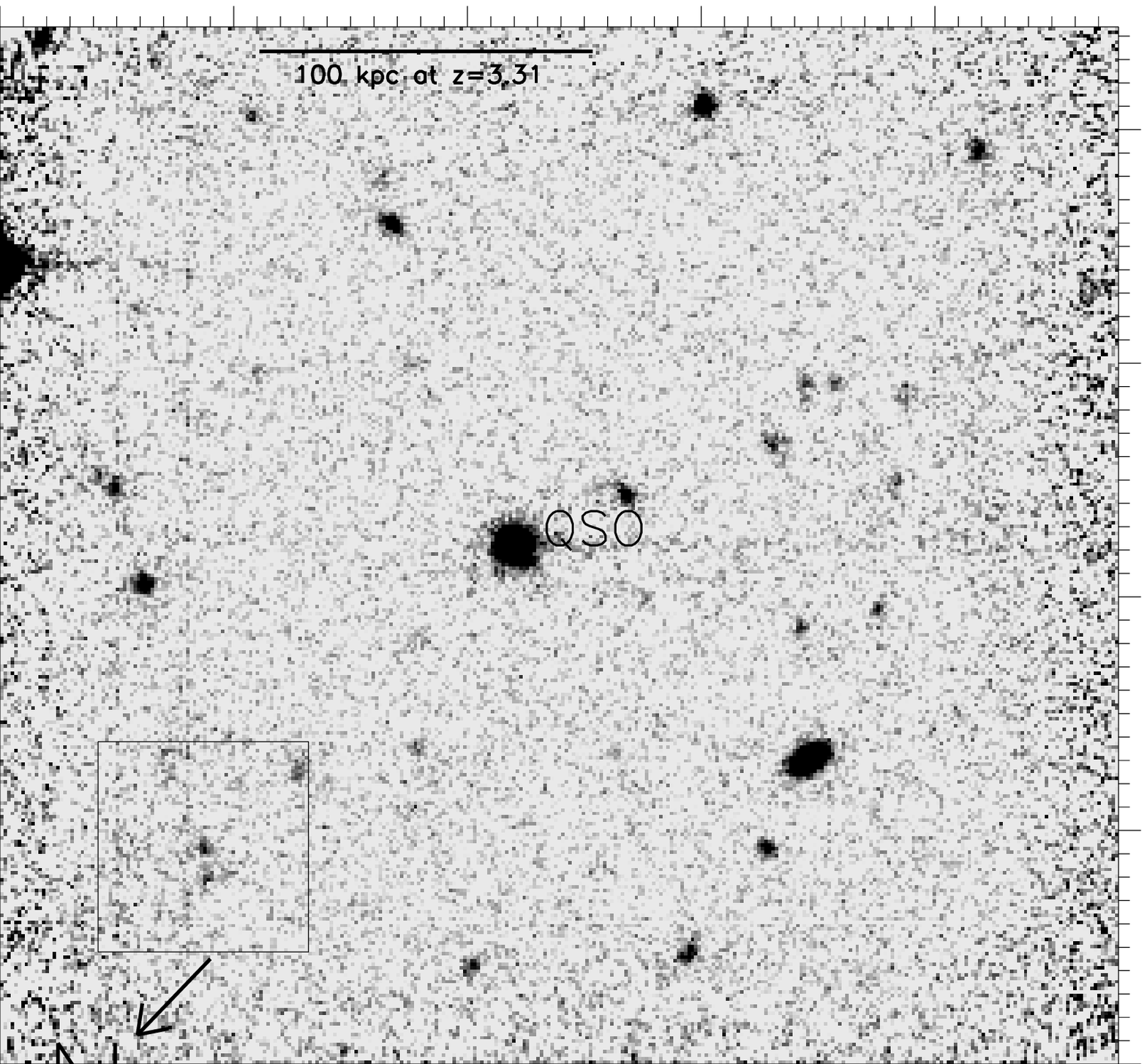}}
\vskip 1in
\hspace{1.5in}
\parbox{1in}{\epsfysize=1in \epsfbox{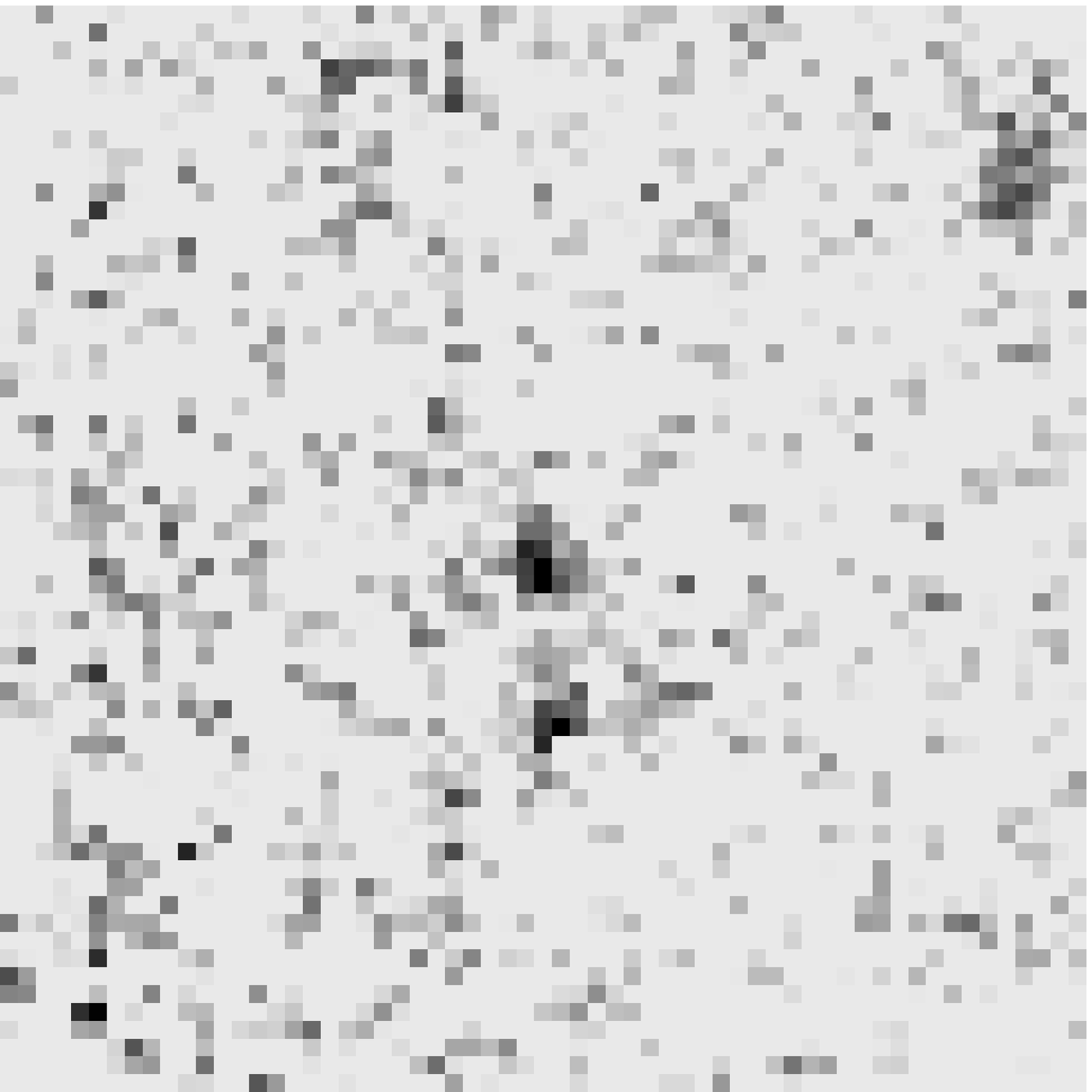}}
\hspace{0.5in}
\parbox{1in}{\epsfysize=1in \epsfbox{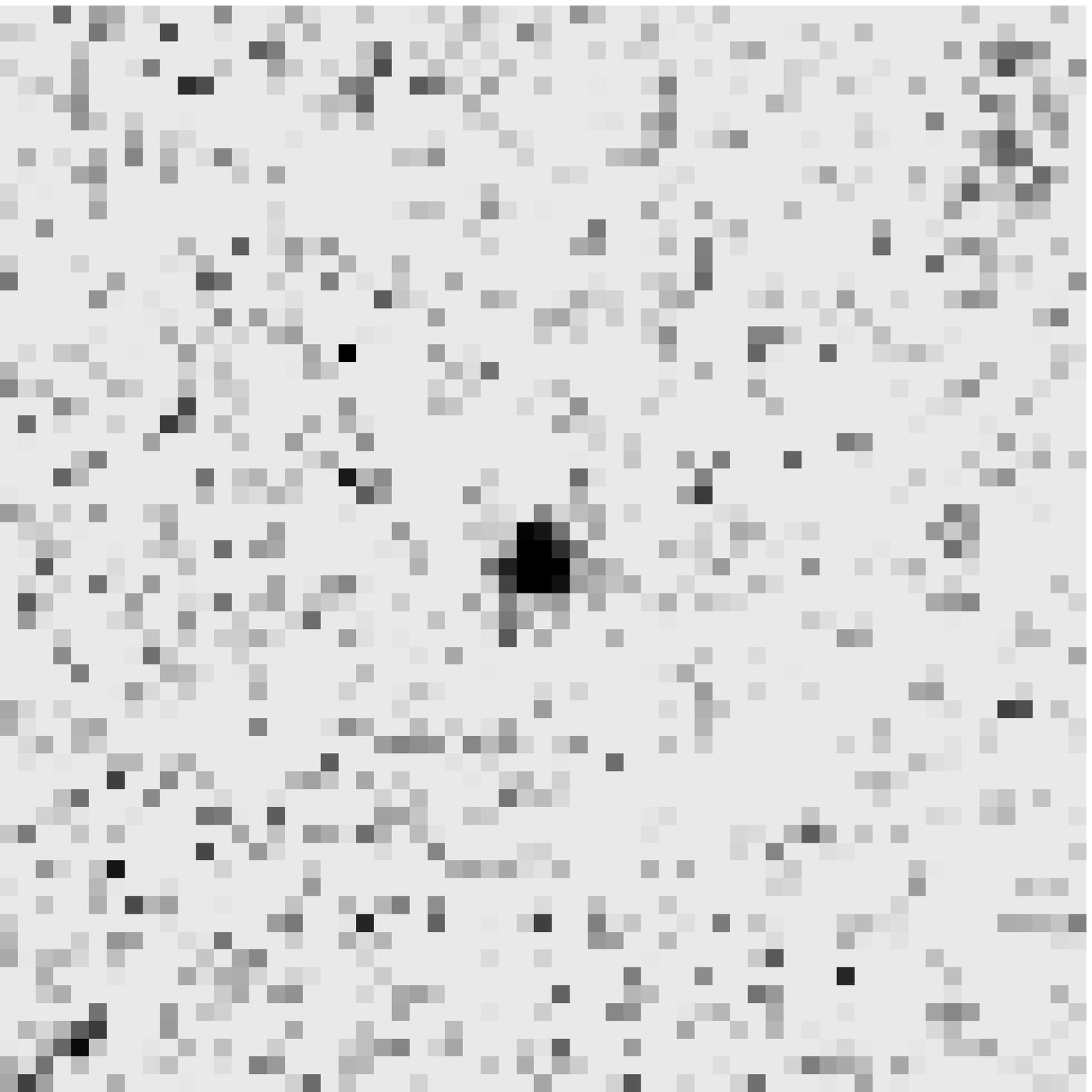}}
\end{figure}

\clearpage

\begin{figure}
\plotone{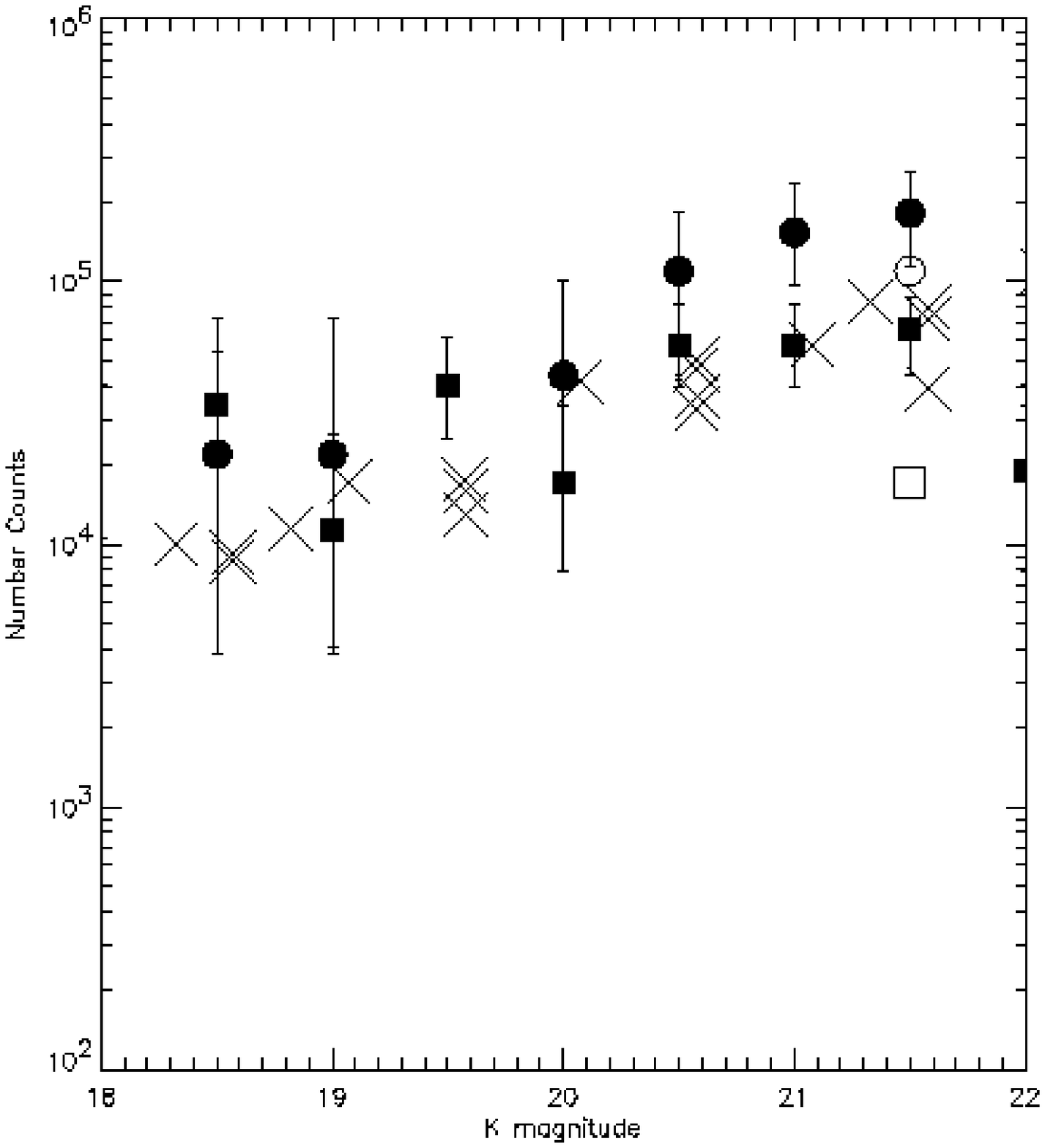}
\end{figure}

\end{document}